# ADPCM WITH NONLINEAR PREDICTION

*Marcos Faúndez-Zanuy\*, Oscar Oliva-Suarez\*\**

\*Escola Universitària Politècnica de Mataró
\*\*Signal Theory & Communications Department (UPC)
Avda. Puig i Cadafalch 101-111, E-08303 Mataró (BARCELONA)
tel:+34 3 757 44 04 fax:+34 3 757 05 24
e-mail: faundez@eupmt.es http://www.eupmt.es/veu

**ABSTRACT**

Many speech coders are based on linear prediction coding (LPC), nevertheless with LPC is not possible to model the nonlinearities present in the speech signal. Because of this there is a growing interest for nonlinear techniques. In this paper we discuss ADPCM schemes with a nonlinear predictor based on neural nets, which yields an increase of 1-2.5dB in the SEGSNR over classical methods. This paper will discuss the block-adaptive and sample-adaptive predictions.

**1. INTRODUCTION**

Mumolo et alt. ([1]) proposed an ADPCM with nonlinear prediction based on Volterra Series, which has the problem of unstability. In [2] we studied a nonlinear prediction model based on neural nets because they achieve higher prediction gains than Volterra (even with a smaller number of coefficients) and are always stable (see [3]). In [4] we proposed an ADPCM scheme with nonlinear prediction (the same model studied in [2]) and a novel hybrid ADPCM-Backward scheme which combines linear and nonlinear prediction (see fig. 1) in order to achieve always the greatest prediction gain. Although the averaged SNR is greater for the nonlinear predictor than for the linear predictor, in several frames the LPC outperforms the NLPC. In [4] we also include an exhaustive study about the training parameter set for having a good generalization capability (and thus robustness over parameter quantization and/or input signals' perturbations). Our final results are between 1 and 2.5 dB over the classical LPC-10 for the range between 2 and 5 quantization bits, while Mumolo is 1dB over LPC-10 for 3 and 4 quantization bits. (We achieve higher gains and wider bit rates).

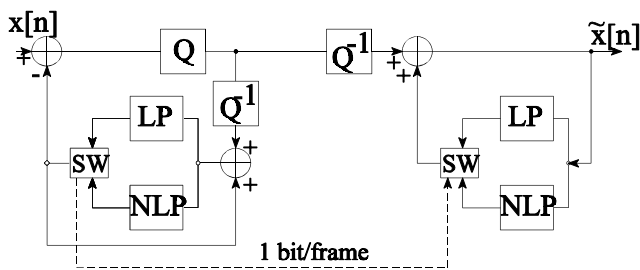

Fig. 1 ADPCM-B hybrid coder. LP: linear predictor, NLP: nonlinear predictor, SW: switch

In [5] the ADPCM forward with NLPC scheme that was proposed obtained similar performance than the LPC one but with one bit less in the quantizer. Thus, for instance the 40Kbps ADPCM-forward LPC is equivalent (similar SEGSNR) than the ADPCM-forward NLPC at 32Kbps. This implies that for the same bit rate, the NLP outperforms the LPC with 2.5 to 3 dB in SEGSNR (except for a quantizer of only 2 bits) for a wide range of frame length. Unfortunately, this improvement is reduced in the backward configuration, due to the different conditions between training and testing (in the backward configuration the neural net (NN) weights are computed over a different frame than the frame they are used for prediction, so the increase in SEGSNR of the NLPC over the LPC is reduced to 2 dB). In order to increase the performance of the backward configuration, we propose to optimize the frame length and to compute the NN weights more frequently. In the limit, if one new neural net is computed for each sample of the signal to encode, we obtain a sample adaptive ADPCM with nonlinear prediction scheme.

In this paper we study the influence of the frame length in the system performance, with the main objective of proposing a sample adaptive ADPCM-Backward Nonlinear prediction speech coder.

**2. ADPCM WITH NONLINEAR PREDICTOR SCHEME**

In order to compare the nonlinear speech prediction system, ADPCM waveform coder is used. The nonlinear predictor is compared against the traditional LPC one, with the following characteristics:

**System overview**
*Predictor coefficients updating*
- The coefficients are updated once time every frame.
- To avoid the transmission of the predictor coefficients an ADPCM backward (ADPCMB) configuration is adopted. That is, the coefficients of the predictor are computed over the decoded previous frame, because it is already available at the receiver and it can compute the same coefficient values without any additional information. The obtained results with a forward unquantized predictor coefficients (ADPCMF) are also provided for comparison purposes.
- The nonlinear analysis consists on a multilayer perceptron with 10 input neurons, 2 hidden neurons and 1 output neuron. The network is trained with the Levenberg-Marquardt algorithm.
- The linear prediction analysis of each frame consists on an all-pole filter, 10 coefficients

obtained with the autocorrelation method (LPC-10) and 25 order filter (LPC-25).

*Residual prediction error quantization*
 ! The prediction error has been quantized with 2 to 5 bits. (bit rate from 16Kbps to 40Kbps).
 ! The quantizer step is adapted with multiplier factors, obtained from [6]. $\Delta_{max}$ and $\Delta_{min}$ are set empirically.

*Database*
 ! The results have been obtained with the following database: 8 speakers (4 males & 4 females) sampled at 8Khz and quantized at 12 bits/sample.

Additional details about the predictor and the database were reported in [2].

*Adpcm Backward- Hybrid Waveform Coder*
In [4] we proposed an ADPCM-Backward hybrid waveform coder with a linear/non linear switched predictor in order to choose always the best predictor and to increase the SEGSNR of the decoded signal. For each frame the outputs of the linear and nonlinear predictor are computed simultaneously with the coefficients computed from the previous encoded frame. Then a logical decision is made that chooses the output with smaller prediction error. This implies an overhead of 1 bit for each frame that represents only 1/200 bits more per sample if the frame size is 200 samples. Obviously, a sample hybrid adaptive scheme is not possible if the decision bit is transmitted to the decoder, so if it is desired to keep the structure of figure 1, the switched decision must be made over transmitted information in order to avoid the transmission of 1 bit/sample for this decision. In this paper, this scheme is not studied due to the high computational burden of a good decision predictor plus LPC and NLPC models computation for each sample of the speech signal to be encoded.

## 3. SAMPLE ADAPTIVE ADPCM-BACKWARD SCHEME

In order to improve the performance of the coder, the frame length can be reduced, because then the coefficients of the predictor are actualized more frequently. The results have been evaluated using subjective criteria (listening to the original and decoded files), and objective criteria, measuring the SEGSNR.
Table 1 shows the SEGSNR obtained with the ADPCM Block adaptive configuration for the whole database with the following predictors: LPC-10, LPC-25 and MLP 10x2x1, with a frame length of 100 samples. Comparing this table with the one presented in [4] it is possible to observe that the SEGSNR has been increased in 1.5 dB for the NLPC and only 0.5 dB in the LPC. These results reveal the superiority of the nonlinear predictor in the forward configuration (2dB aprox. over LPC-25 except for the 2 bit quantizer). This superiority is greater if the quantizer has a high number of levels.
In the backward configuration there is a small SEGSNR decrease with the linear predictor versus the forward configuration. For the nonlinear predictor it is more significative (nearly 3dB), but the SEGSNR is better than LPC-10 except for Nq=2 bits. Also, the variance of the SEGSNR is greater than for the linear predictor, because in the stationary portions of speech the neural net works satisfactorily well, and for the unvoiced parts the nnet generalizes poorly. For this reason, a hybrid predictor is proposed in [4]. Also from the figures of [4] and [5] that show the evolution of the SEGSNR as function of the frame length, it can be seen that if the frame length is reduced under 50 samples, the SEGSNR falls drastically. Therefore, we propose the following procedure for obtaining and efficient sample adaptive scheme:
a) We define a training window, which defines the number of samples used for computing the LPC and NLPC coefficients.
b) We define the actualization rate of the coefficients, which is equivalent to a computing window, defining the number of samples for which the same predictor coefficients are used. Obviously, if the length of the computing window is 1 sample, the scheme is sample adaptive, instead of block-adaptive.
The results of the ADPCM forward (with unquantized predictor coefficients) are also provided such us reference of the backward configuration.

| METHOD | Nq=2 bits | | Nq=3 bits | | Nq=4 bits | | Nq=5 bits | |
|---|---|---|---|---|---|---|---|---|
| | Segsnr | std | Segsnr | std | Segsnr | std | Segsnr | std |
| ADPCMFLPC10 | 15.35 | 5.8 | 21.18 | 6.4 | 25.86 | 6.9 | 30.52 | 7.1 |
| ADPCMFLPC25 | 15.65 | 5.6 | 21.46 | 6.4 | 26.26 | 6.9 | 30.79 | 7.2 |
| ADPCMFMLP | 15.5 | 7.4 | 24.12 | 7.3 | 29.35 | 7.6 | 34.14 | 8.4 |
| ADPCMBLPC10 | 14.92 | 5.1 | 20.59 | 5.9 | 25.38 | 6.6 | 30.02 | 7.1 |
| ADPCMBLPC25 | 14.88 | 5.1 | 20.95 | 5.5 | 25.2 | 6 | 30.1 | 6.2 |
| ADPCMBMLP | 14.35 | 6.9 | 21.48 | 7.5 | 26.76 | 7.6 | 31.5 | 8.4 |
| ADPCMB-HYBRID | 16.1 | 4.8 | 22.38 | 5.8 | 27.51 | 6.1 | 32.53 | 6.4 |

Table 1. SEGSNR for ADPCM forward, backward, linear, nonlinear and hybrid.

### 3.1 Efficient initialization algorithm
Obviously this scheme requires a high number of computations, so we must propose first an efficient algorithm for training the neural nets with a reduced complexity.
In the original training algorithm [4] a multistart algorithm is used, which consists on computing several random initializations (experimentally fixed in [4] to 4). And 6 epochs for initialization (fixed in [4]). I must be taken into account that the increase in the computational burden is limited because of the good convergence speed of the Levenberg-

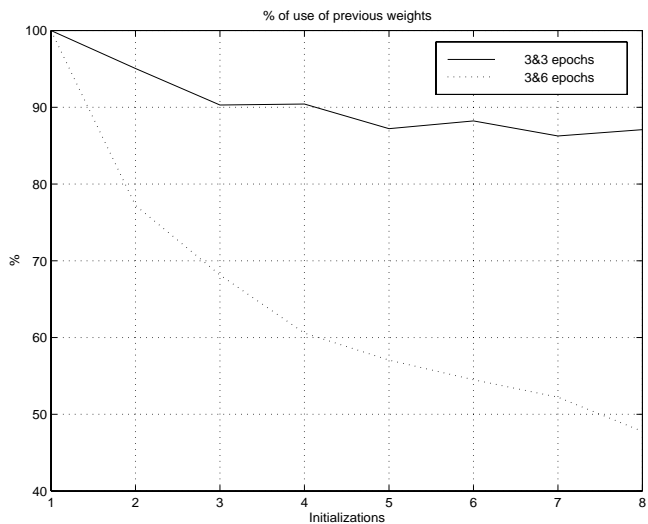

Fig. 2 selection of previous weights vs number of random initializations

Marquardt algorithm. After this process, the initialization that gives the higher SEGSNR over the training frame is chosen. This method yields good results, but frequently, consecutive frames are very similar, so their associated weights will also be similar. For this reason, we propose to change one of the random initializations for a deterministic initialization: the weights of the previous frame.

We have made an exhaustive study and we have observed that this initialization does not need to be as much trained as the random initialization, because it is near the optimal solution, so this initialization is only trained during 3 epochs. Figure 2 shows the percentage of selection of this deterministic initialization, in two cases:
a) The random initializations are trained 3 epochs.
b) The random initializations are trained 6 epochs.
In order to compare the improvements in speed, figure 3 represents the number of samples processed per second, in a Pentium 166 MMX. Figure 4 represents the SEGSNR vs the number of adaptive quantization bits using previous weights (3 epochs) plus one random initialization (6 epochs) compared against four random initialization (6 epochs).

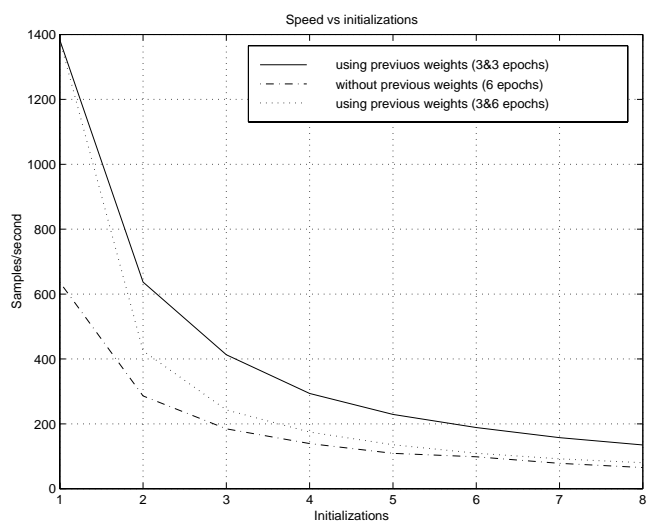

Fig. 3 Samples per second vs number of random initializations

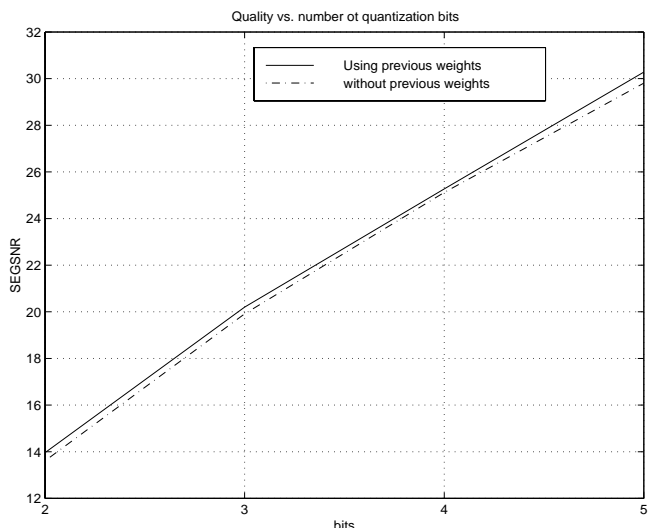

Fig. 4 SEGSNR vs number of quantization bits.

In figures 2,3 and 4 it can be seen that if one of the weights' initializations is equal to the weights of the previous frame, then more speed and higher SEGSNR can be achieved over the baseline algorithm proposed in [4], so this novel initialization algorithm will be considered in the other experiments. Oviously, this modification can be used in two ways: to increase the SEGSNR at the same computational complexity, and to reduce the computational complexity achieving the same SEGSNR.

### 3.2 Sample adaptive ADPCM-Backward NLPC scheme
With the main goal of propose a sample adaptive scheme, the influence of two parameters is studied:
a) The training window.
b) The computing window.
Figures 5, 6 and 7 let us to explain the main conclusions: Although the behaviour of these figures does not show a clear tendency, it seems that if the computing window is short, the performance of the system degrades. Also, the computational complexity is increased, because 4 neural nets are trained for each sample of the speech signal, and this is computational expensive.
Altougth we don't understand very well the degradation in SEGSNR for small computing windows, be belive that it is due to a problem with the quantization step adaptation: the predictor is changed too frequently, giving different error levels from the previous predictor, and the quantizer is unable to track this changes, so really there is a bad step adaptation. Keep in mind that the multiplying factors obtained from [6] were computed in different conditions than the conditions of this experiment, so it is not assured that it works properly unless it would be recalculated under more realistic conditions.
Another relevant fact is that in the linear case, the block adaptive solution can achieve much better prediction (see

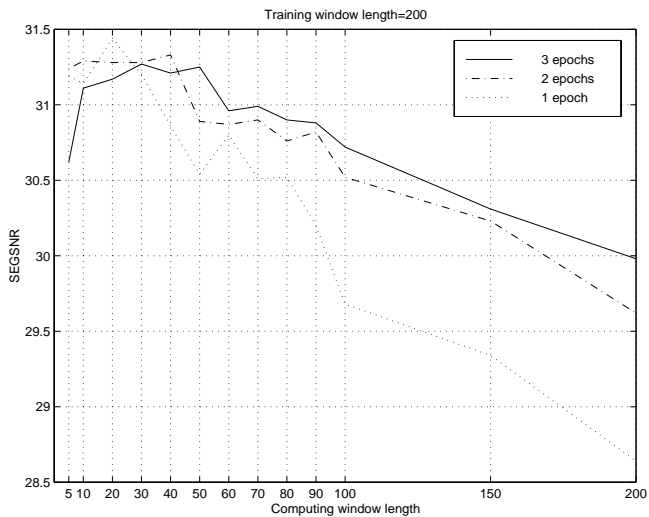

Fig. 5 SEGSNR for training window of 200 samples

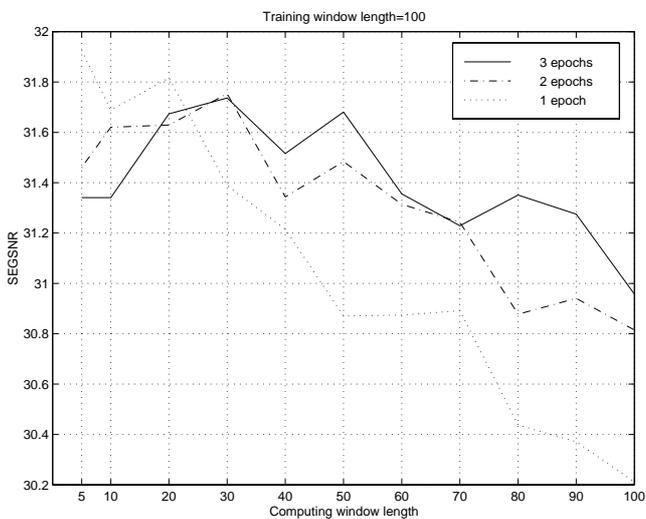

Fig. 6 SEGSNR for training window of 100 samples

[7,pp.229]) than a sample adaptation with gradient methods (LMS or LMA) for a high order all-pole filter. For this reason, the LPC sample adaptive scheme is not included.

## 4. COMPARISON WITH PREVIOUSLY PUBLISHED WORK

The unique work that we have found that deals with ADPCM with nonlinear prediction is the one proposed by Mumolo et alt. [3]. It has problems of unstability, which were overcome with a switched linear/nonlinear predictor.

Our novel nonlinear scheme has been always stable in our experiments.

The results of our novel scheme show an increase of 1 to 2.5 dB over classical LPC-10 for quantizer ranges from 2 to 5 bits, while the work of Mumolo [3] is 1 dB over classical LPC for quantizer ranges from 3 to 4 bits and also with and hybrid predictor.

The improvement can be increased if the frame length is decreased to an appropriate value, at the cost of more computational complexity.

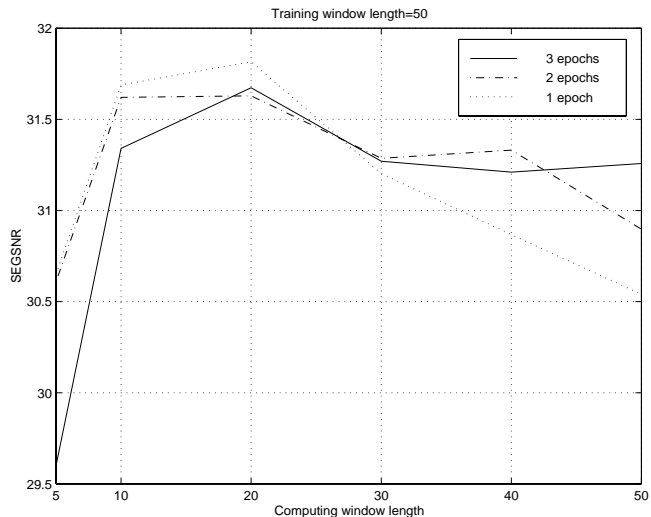

Fig. 7 SEGSNR for training window of 50 samples

## ACKNOWLEDGEMENTS
This work has been supported by the CICYT TIC97-1001-C02-02